\documentclass[conference,10pt, a4paper]{IEEEtran}
\usepackage{amsmath}
\usepackage{amssymb}
\usepackage{xcolor}
\usepackage{amsthm}
\usepackage[normalem]{ulem}
\usepackage[font=normalsize]{caption}
\usepackage{siunitx}
\usepackage{cuted}
\usepackage{tabularx}
\usepackage{multirow}
\usepackage{booktabs}
\usepackage{diagbox}
\usepackage{subcaption}
\usepackage[inline]{enumitem}
\usepackage[boxruled,norelsize]{algorithm2e}
\makeatletter
\newcommand{\removelatexerror} {\let\@latex@error\@gobble}
\makeatother

\usepackage{tcolorbox}
\tcbuselibrary{many}

\newcommand{\highlightcolor}{black}

\newcommand{\superscript}[1]{^{\text{#1}}}

\newcommand{\highlight}[1]{{\textcolor{\highlightcolor}{#1}}}

\theoremstyle{plain}% from 'amsthm'
\tcolorboxenvironment{theorem}{
	enhanced jigsaw, colframe=black, interior hidden,
	breakable,before skip=10pt,after skip=10pt}

\tcolorboxenvironment{lemma}{
	enhanced jigsaw, colframe=black, interior hidden,
	breakable,before skip=10pt,after skip=10pt}

\tcolorboxenvironment{corollary}{
	enhanced jigsaw, colframe=black, interior hidden,
	breakable,before skip=10pt,after skip=10pt}

\newtcbtheorem[]{alg}{Algorithm}{fonttitle=\bfseries}{alg}

\DeclareSIUnit\bit{bits}

\begin{document}

\title{Recursive Optimization of Finite Blocklength Allocation to Mitigate Age-of-Information Outage}

% Censor authors and affliations for the double-blind review

\author{
	\IEEEauthorblockN{Bin~Han\IEEEauthorrefmark{1}, Zhiyuan~Jiang\IEEEauthorrefmark{2},Yao~Zhu\IEEEauthorrefmark{3}, %Yulin~Hu\IEEEauthorrefmark{3},
	and Hans~D.~Schotten\IEEEauthorrefmark{1}\IEEEauthorrefmark{5}}\\
	\IEEEauthorblockA{
	\IEEEauthorrefmark{1}University~of~Kaiserslautern,~%
	\IEEEauthorrefmark{2}Shanghai~University,~%
	\IEEEauthorrefmark{3}RWTH~Aachen~University,\\%
	\IEEEauthorrefmark{5}German Research Center for Artificial Intelligence
	}%
}

% The paper headers
%\markboth{Journal of \LaTeX\ Class Files,~Vol.~14, No.~8, August~2015}%
%{Shell \MakeLowercase{\textit{et al.}}: Bare Demo of IEEEtran.cls for IEEE Journals}

% make the title area
\maketitle

% As a general rule, do not put math, special symbols or citations
% in the abstract or keywords.
\begin{abstract}
\highlight{As an emerging metric for the timeliness of information delivery, Age-of-Information (AoI) raises a special interest in the research area of tolerance-critical communications, wherein sufficiently short blocklength is usually adopted as an essential requirement. However, the interplay between AoI and finite blocklength is scantly treated. This paper studies the occurrence of critically high AoI, i.e., AoI outage, in generic orthogonal multi-access (OMA) systems with respect to the blocklength allocation among users. A Markov Decision Process model is set up for the problem, which enables a steady-state analysis. Therewith a recursive optimizer to reduce AoI outage is proposed, and further enhanced by heuristic penalty functions. The burstiness of AoI outage is also analyzed to provide additional insights into this problem in the finite blocklength (FBL) regime. It is shown that, different from average AoI optimizations, a risk-sensitive approach is significantly beneficial for AoI outage optimizations, on account of the FBL regime. }
\end{abstract}

% Note that keywords are not normally used for peerreview papers.

\IEEEpeerreviewmaketitle

\section{Introduction}\label{sec:intro}
\highlight{In context of future beyond-5G (B5G) and 6G wireless networks, especially beyond the Ultra-Reliable Low Latency Communication (URLLC) scenario}, the timeliness of information delivery is becoming more and more important to guarantee reliable functionality for tolerance critical applications such as factory automation and remote control. In the past years, the Age-of-Information (AoI) \cite{KYG2012real} concept has been intensively investigated as a practical metric for information timeliness. It basically captures the time elapsed since generation time of the latest information at an intended terminal. Recent studies have found that, in many wireless control applications such as cooperative autonomous driving, an AoI optimizazion is more efficient than simple communication latency minimization. 

Therefore, a series of studies have been conducted, focusing on AoI optimizations. More specifically, the retransmission scheme~\cite{HLS+2019real} and the packet scheduling problem~\cite{Hsu2018age,KSU+2018scheduling,JZNC2019unified,JKZ+2018decentralized,JKZN2018can} towards reduced AoI in communication networks. Recently, we have reported a pioneering work that focus in the finite blocklength (FBL) regime and considers the AoI minimization as a radio resource allocation problem~\cite{HZJ+2019optimal}. 

In many practical scenarios \highlight{such as autonomous driving and factory automation,} the application is somewhat predictive on higher layers, and therefore exhibits tolerance to information obsolescence at a certain degree. In this context, it invokes an investigation on the robustness of communication links in perspective of information timeliness. More specifically, we are interested in the events of AoI outage, i.e. when the AoI in a communication system exceeds some threshold. \highlight{In this paper, we take a perspective of radio resource allocation to study this problem in the FBL regime}.

\highlight{Our main contributions presented in this paper consists of three parts. First, we have built a statistical model of AoI outage in orthogonal multi-access (OMA) FBL communication systems, and have proved it to rely on the steady-state distribution of the system, which is uniquely determined by the blocklength allocation policy. Second, we have proposed a recursive framework to optimize the blocklength allocation towards an outage rate minimization, which is proven efficient when deployed with a heuristic penalty function. Additionally, concerning the FBL-specific risk of lasting outages, we have proposed an analytical model for the bursty features of AoI outages, which accurately fits our simulation results.}

The remaining content of this paper is organized as follows. First, in Sec.~\ref{sec:related_work} we address some existing works that are relevant to this study in different perspectives. Then we set up the Markov AoI model in Sec.~\ref{sec:model}, and in Sec.~\ref{sec:ssa_pi} we propose a recursive policy optimizer based on steady-state analysis of the model. As a complement we also analyze the burstiness of AoI outages under arbitrary blocklength allocation policy in Sec.~\ref{sec:burstiness}. After reporting in Sec.~\ref{sec:sim} the simulations to demonstrate our proposed approaches, we conclude this paper with Sec.~\ref{sec:concl} and provide a few outlooks to future research.

\section{Related Work}\label{sec:related_work}
URLLC has been addressed as a characterizing feature of 5G wireless networks~\cite{ITU2017minimum,3GPP2017tr38802}. Aiming at emerging tolerant-critical applications -- such as remote control, factory automation, autonomous driving, etc. -- it requires a link reliability above $99.999\%$ with a latency below \SI{10}{\milli\second}. To cope with such strict performance requirements, significant efforts have been made from various perspectives. This includes new radio frame design with shorter transmission time interval (TTI)~\cite{PKBF2017rethink}, grant-free access scheduling scheme~\cite{EBDG2019radio}, and exploitation of resource diversity~\cite{KNBP2018uplink,RPD+2019resource}. These studies generally focus on the classical concept of latency. As a complement to them, in the past few years, a new interest has been triggered to investigate the age of information in communication systems. 
	
The concept of AoI describes the difference between the current time and the latest updated (successfully received) information. Compared to the classical concept of latency, it is more efficient and intuitive in describing the timeliness of information delivery, while also presenting more complex behavior. Most existing works on AoI optimizations in wireless networks have been focused on scheduling of terminals. Specifically, when the scheduling decisions are centralized, correspondingly a broadcast network scenario, optimal (or near-optimal) scheduling policies are derived considering various channel conditions and frame structures \cite{Hsu2018age,KSU+2018scheduling}. On the other hand, decentralized medium access in wireless uplinks requires the scheduling decisions to be distributed. Therefore, considerable amount of work has been dedicated to this topic \cite{JZNC2019unified,JKZ+2018decentralized,JKZN2018can}. However, few works have considered the impact of finite blocklength on AoI, whereas in order to ensure low AoI and to deliver small packets for status update, a short blocklength is usually essential. Moreover, on account of the AoI optimizations with finite blocklength, few existing works, except related literature on risk-sensitive optimizations \cite{GSKN2018risk}, have studied the issue of sparse reward (cost), which results from the unique nature of finite blocklength transmissions since outage, although rarely occurs, is of critical importance. 

In the finite blocklength (FBL) regime, the data transmission is no longer arbitrarily reliable,especially when the blocklength is short. In this regime, the actual error probability becomes significantly higher than the Shannon limit, which assumes the blocklength to be infinite or sufficiently long. To tackle this problem, \emph{Polyanskiy} et al. have derived an accurate approximation  of the achievable coding rate within FBL regime for AWGN channels~\cite{Poly_2010_FBL}. Subsequently, the performance of FBL in variant scenarios have been studied, including wireless power delivery~\cite{Khan_2017_FBL}, secure communications~\cite{Wang_2019_FBL}, and relaying networks~\cite{Hu_2019_FBL}. To further improve the reliability, retransmission schemes are also revisited with FBL code~\cite{Zhu_2019_FBL}~\cite{Avranas_2019_FBL}. Despite of these progresses, most existing FBL works only focus on the end-to-end delay, leaving the topic of AoI in FBL communication systems at the research frontier.

Looking forward to bridge the gap between aforementioned areas, in the following sections we investigate the problem of finite blocklength allocation in \highlight{generic OMA} systems, by means of minimizing the rate of AoI outages, i.e. when at least one device exceeds an AoI-threshold.

\section{System Model}\label{sec:model}
We consider the two-device AoI problem in perspective of PHY-layer transmission errors, where each of two devices $m\in\{1,2\}$ periodically generates an up-to-date message of package length $d$ (active source), and send it to a Multi-Access Edge Computing (MEC) server. \highlight{The uplink transmissions are executed in an OMA scheme, e.g. Orthogonal Frequency-Division Multi-Access (OFDMA) or Time-Division Multi-Access (TDMA), where both devices are synchronous to the same time frame, and the two messages in every frame share a finite blocklength $N$ for transmission}. If a message fails to be delivered to the server, no retransmission is scheduled, and the device will just transmit the latest message in every period. For simplification we assume the server as always idle to process any incoming message, and the uplink channels of both devices as Gaussian channels under independent Bernoulli block fading, and their channel state information (CSI) are fully available at the MEC server. More specifically, the SNR of device $m\in\{1,2\}$ in the $k^\text{th}$ transmission period is 
\begin{equation}
	\gamma_{k,m}=\begin{cases}
		\gamma_\text{good}&\text{if $x_{k,m}=1$};\\
		\gamma_\text{bad}&\text{otherwise},
	\end{cases}\label{eq:bernoulli_channel}
\end{equation}
and $x_{k,m}$ is the $k^\text{th}$ sample of Bernoulli process $X_m\sim \mathcal{B}(\alpha_m)$ where $\alpha_m$ is the chance of device $m$ having a good channel.

Upon the success/failure of transmission, the AoI for device $m$ at the end of $k\superscript{th}$ period is
\begin{equation}
A_{k,m}=\begin{cases}
1&\text{success};\\
A_{k-1,m}+1&\text{failure}.
\end{cases}
\end{equation}
The transmission failure rate $\varepsilon$ in FBL regime is approximately
\begin{equation}\label{eq:err_rate}
	\varepsilon_{k,m}\approx Q\left(\sqrt{\frac{n_{k,m}}{V_{k,m}}}\left(\mathcal{C}_{k,m}-\frac{d}{n_{k,m}}\right)\ln 2\right),
\end{equation}
\highlight{where $n_{k,m}$ is the blocklength allocated to device $m$, so that $n_{k,1}+n_{k,2} \equiv N$.} $V_{k,m}$ and $\mathcal{C}_{k,m}$ are the dispersion and Shannon capacity of the UL channel for device $m$ in period $k$, respectively. Especially, for Gaussian channels
\begin{align}
	V_{k,m}&=1-\frac{1}{(1+\gamma_{k,m})^2},\label{eq:dispersion}\\
	\mathcal{C}_{k,m}&=B_{k,m}\log_2(1+\gamma_{k,m}),\label{eq:capacity}
\end{align}
where $B_{k,m}$ is the UL channel bandwidth for device $m$ in period $k$. For simplification we consider a normalized bandwidth $B_{k,1}=B_{k,2}=B=\SI{1}{\hertz}$ for all $k\in\mathbb{N}$, as commonly done in the field of FBL information theory.

Now we investigate a consistent scheduling policy where the blocklength allocation in every period $\mathbf{n}_k=(n_{k,1},n_{k,2})$ is determined by the previous system state vector $\phi_{k-1}=(A_{k-1,1},A_{k-1,2},x_{k-1,1},x_{k-1,2})$ with respect to a consistent allocation policy $\mathbf{n}_k=\mathcal{L}(\phi_{k-1})$. Then the system state becomes an infinite-state Markov chain

\begin{strip}
	\begin{tcolorbox}
		\begin{equation}
		\footnotesize
		\text{Prob}(\phi_{k}\vert\phi_{k-1})=
		\begin{cases}
		\varepsilon_{k,1}(\phi_{k-1})\varepsilon_{k,2}(\phi_{k-1})P_{X_1}(x_{k,1})P_{X_2}(x_{k,2})&\phi_k=(A_{k-1,1}+1,A_{k-1,2}+1,x_{k,1},x_{k,2})\\
		\varepsilon_{k,1}(\phi_{k-1})[1-\varepsilon_{k,2}(\phi_{k-1})]P_{X_1}(x_{k,1})P_{X_2}(x_{k,2})&\phi_k=(A_{k-1,1}+1,1,x_{k,1},x_{k,2})\\
		[1-\varepsilon_{k,1}(\phi_{k-1})]\varepsilon_{k,2}(\phi_{k-1})P_{X_1}(x_{k,1})P_{X_2}(x_{k,2})&\phi_k=(1,A_{k-1,2}+1,x_{k,1},x_{k,2})\\
		[1-\varepsilon_{k,1}(\phi_{k-1})][1-\varepsilon_{k,2}(\phi_{k-1})]P_{X_1}(x_{k,1})P_{X_2}(x_{k,2})&\phi_k=(1,1,x_{k,1},x_{k,2})\\
		0&\text{otherwise}
		\end{cases}\label{eq:transition}
		\end{equation}
	\end{tcolorbox}
\end{strip}

\highlight{As indicated in} \cite{HZJ+2019optimal}, there is always a value $A_\text{max}$ that $\text{Prob}(A_{k,m}>A_\text{max})\approx 0$ for all $(k,m)$, which allows us to approximate the system with a finite-state Markov chain. For the convenience of indication, we refer to its state space
\begin{equation}
	\mathfrak{S}=\{1,2,\dots,A_\text{max}\}^2\times\{0,1\}^2=\{\mathbf{s}_1,\mathbf{s}_2,\dots,\mathbf{s}_{4A_\text{max}^2}\},
\end{equation}
where the index $i$ for $\mathbf{s}_i=(s_{i,1},s_{i,2},s_{i,3},s_{i,4})\in\mathfrak{S}$ is set by
\begin{equation}
	i=2\times\{2\times[(s_{i,1}-1)A_\text{max}+s_{i,2}-1]+s_{i,3}\}+s_{i,4}+1.
\end{equation}

This finite-state Markov chain can be thus fully described by the state transition matrix $\mathbf{P}$ where $P_{i,j}=\text{Prob}\left(\phi_{k+1}=\mathbf{s}_j\vert \phi_k=\mathbf{s}_i\right)$. 
Without loss of generality we assume that $\phi_0=\mathbf{s}_1=(1,1,0,0)$, then we can derive the probability that the AoI is a specific value $i$ after $k$ periods:
\begin{equation}\label{eq:outage_prob_from_transition}
	\text{Prob}\left(\phi_k=\mathbf{s}_i\right)=\left(\mathbf{P}^k\right)_{1,i}.
\end{equation}

For reliable timely communication, we are interested in the outages where the AoI of at least one device exceeds some threshold $A_{\text{out}}\le A_\text{max}$, which can be presented as
\begin{equation}
	\begin{split}
		P_\text{out}=&\text{Prob}\left(\exists m\in\{1,2\}: A_m>A_\text{out}\right)
		%\\=&1-\text{Prob}\left(A_m\le A_\text{out}, \forall m\in\{1,2\}\right)
		.
	\end{split}
\end{equation}

\section{\highlight{Steady-State Analysis and Policy Optimizer}}\label{sec:ssa_pi}
%Nowadays it is popular to invoke Reinforcement Learning (RL) techniques such as Q-Learning to solve Markov Decision Processes. RL can also be applied for AoI optimization, for which our previous work~\cite{HZJ+2019optimal} provides an example. Nevertheless, when setting the outage probability $P_\text{out}$ as the objective function to minimize, especially in the context of reliable communications, the applicability of sampling-based RL approaches can be challenged. More specifically, a straightforward design of the reward function for a RL method to solve our problem here is binary:
\highlight{Now we can see that the blocklength allocation in such a system is a Markov Decision Process (MDP) where the transition probabilities between states are statistically determined by the policy of action (i.e. blocklength allocation), so its performance can be optimized w.r.t. the policy. More specifically, aiming at reducing the events of AoI outage, we can design a straightforward binary penalty function of arbitrary state $\phi$:
\begin{equation}
	\zeta_\text{simple}(\phi)=\begin{cases}
		1& \phi\in\mathfrak{S}_\text{out},\\
		0&\text{otherwise},
	\end{cases}
	\label{eq:simple_reward}
\end{equation}
where $\mathfrak{S}_{\text{out}}\subseteq\mathfrak{S}=\left\{\left.\mathbf{s}_i \right\vert \exists m\in\{1,2\}: s_{i,m}>A_{\text{out}}\right\}$, and the we look for an allocation policy $\mathfrak{S}\overset{\mathcal{L}}{\to}[0,1,\dots,N]^2$ that
\begin{equation}
	\begin{split}
		&\min\limits_{\mathcal{L}}~\mathbb{E}\left\{\zeta_\text{simple}(\phi)\right\}=\min\limits_{\mathcal{L}}~\sum\limits_{\mathbf{s}\in\mathfrak{S}}\zeta_\text{simple}(\phi)\text{Prob}(\phi=\mathbf{s})\\
		=&\min\limits_{\mathcal{L}}\sum\limits_{\mathbf{s}\in\mathfrak{S}_\text{out}}\text{Prob}(\phi=\mathbf{s})=\min\limits_{\mathcal{L}}P_\text{out}.
	\end{split}
\end{equation}}
%However, in the context of reliable communications, $P_\text{out}$ is usually on such a low level that it takes hundreds or even more samplings to obtain the first zero-reward even under an average policy. The RL solver, in this case, may converge early to an arbitrary policy, which is known as the sparse reward problem in RL \cite{RHL+2018learning}.  Towards an efficient and reliable policy solver to minimize the AoI outage probability, we rely on the steady-state features of the discussed problem. 

%There are also alternative reinforcement learning methods that are designed to handle high-reliability problems, such as the risk-sensitive approach reported in \cite{GSKN2018risk}. Nevertheless, such methods generally rely on heuristic reward functions to approximate the global optimum, instead of guaranteeing to minimize the outage probability.

\subsection{Steady-State Analysis}
First, it is trivial to derive that the Markov process under discussion is ergodic and recurrent, hence the system has a unique steady-state distribution $\mathbf{\Pi}=(\pi_1,\pi_2,\dots\pi_{4A_{\text{max}}^2})$, where $\pi_i=\lim\limits_{k\to+\infty}\text{Prob}(\phi_k=\mathbf{s}_i)$. Its value can be obtained by solving the linear equation
\begin{equation}\label{eq:steady-state_dist}
	(\mathbf{I}-\mathbf{P}^{\text{T}})\times\mathbf{\Pi}^{\text{T}}=\mathbf{0}.
\end{equation}
Furthermore, it is also trivial to derive that there is some finite interval $k_\text{cvg}$ where the distribution of $\phi_k$ converges to $\mathbf{\Pi}$, i.e.
\begin{equation}\label{eq:transfer_matrix_convergence}
	\text{Prob}(\phi_k=\mathbf{s}_i)\approxeq\pi_i,\forall k\ge k_\text{cvg},\forall \mathbf{s}_i\in\mathfrak{S}.
\end{equation}
And from the ergodicity we know
\begin{equation}\label{eq:ergodicity}
P_\text{out}=\lim\limits_{K\to+\infty}\frac{1}{K}\sum_{k=1}^{K}\sum_{\mathbf{s}\in\mathfrak{S}_{\text{out}}}\text{Prob}\left(\phi_k=\mathbf{s}\right).
\end{equation}

and the outage probability becomes 
\begin{equation}
	\begin{split}
		P_\text{out}=&\lim\limits_{K\to+\infty}\frac{1}{K}\sum_{k=1}^{K}\sum_{\mathbf{s}\in\mathfrak{S}_{\text{out}}}\text{Prob}\left(\phi_k=\mathbf{s}\right)\\
				=&\sum_{\mathbf{s}\in\mathfrak{S}_{\text{out}}}\highlight{\underbrace{\left[\lim\limits_{K\to+\infty}\frac{1}{K}\sum_{k=1}^{k_\text{cvg}}\text{Prob}\left(\phi_k=\mathbf{s}\right)\right.}_{=0}}\\
				&\left.+\lim\limits_{K\to+\infty}\frac{1}{K}\sum_{k=k_\text{cvg}}^{K}\text{Prob}\left(\phi_k=\mathbf{s}\right)\right]\\
				=&\sum_{\mathbf{s}\in\mathfrak{S}_{\text{out}}}\lim\limits_{K\to+\infty}\frac{1}{K}\sum_{k=k_\text{cvg}}^{K}\text{Prob}\left(\phi_k=\mathbf{s}\right)\\
				\overset{\eqref{eq:transfer_matrix_convergence}}{=}&\sum_{i:\mathbf{s}_i\in\mathfrak{S}_{\text{out}}}\lim\limits_{K\to+\infty}\frac{1}{K}\sum_{k=k_\text{cvg}}^{K}\pi_i\\
		=&\sum\limits_{i:\mathbf{s}_i\in\mathfrak{S}_\text{out}}\pi_i\label{eq:outage_prob_from_ssd}
	\end{split}
\end{equation}

\subsection{\highlight{Recursive Policy Optimizer to Minimize Outage Rate}}
Eq.~\eqref{eq:outage_prob_from_ssd} implies a self-evident conclusion that the AoI outage probability is depending not on the dynamic converging process of the Markov system, but only on its steady-state distribution $\mathbf{\Pi}$, \highlight{which is uniquely determined by $\mathbf{P}$. This simple construction of the cost function encourages us to recursively optimize the policy $\mathcal{L}$ with respect to $\mathbf{P}$. }

More specifically, our proposed method is described in Figure \ref{alg:policy_iteration}. Environment specifications are globally set as constants. The \texttt{Main} function stores the block allocation policy in an integer vector $\mathbf{\Lambda}$ that $\lambda_i=n_{k,1}\left(\mathbf{s}_i\right)$. It launches with a random initial policy and therewith call the \texttt{CalcSSD} function to calculate the corresponding steady-state distribution $\mathbf{\Pi}$. Afterwards, \texttt{Main} iterates to update the policy in such a way that it minimizes the AoI outage penalty $\mathbb{E}\{\zeta_\text{simple}\}$ -- which is estimated by the \texttt{Pnt} function -- according to the current $\mathbf{\Pi}$. The function \texttt{PTr} is implemented to calculate the transition probability between two arbitrary states upon a blocklength allocation. The iteration is terminated when a convergence in $\mathbf{\Lambda}$ is detected. 

\highlight{It is worth to remark that the time complexity of this recursive algorithm is mainly determined by the solution of $\mathbf{\Pi}$ in \texttt{CalcSSD}, which counts in a generic form $\mathcal{O}\left(\vert\mathfrak{S}\vert^3\right)$ ($\vert\mathfrak{S}\vert=4A_\text{max}^2$ through our discussion in this paper).}

%It is worth to note here that the cost function \texttt{EstOutage} is practically the expectation of $\left(1-\zeta_\text{simple}\right)$ with respect to the system's steady-state distribution, where $\zeta_\text{simple}$ is the simple binary reward defined earlier in Eq.~\eqref{eq:simple_reward}.

%\begin{alg}{Minimize AoI outage rate with PI}{}{}
\begin{figure}[!hbtp]
	\removelatexerror
	\begin{algorithm}[H]
		\label{alg:policy_iteration}
		\scriptsize
		\DontPrintSemicolon
		\SetKwFunction{FMain}{Main}
		\SetKwFunction{FCalcSSD}{CalcSSD}
		\SetKwFunction{FPTr}{PTr}
		\SetKwFunction{FPnt}{Pnt}
		\SetKwProg{Pn}{Function}{:}{}
		\textbf{Global}: $\gamma_\text{good},\gamma_\text{bad},\alpha_1,\alpha_2,A_\text{max},\mathfrak{S}_\text{out},\epsilon_\text{cvg}$\;
		\;
		\Pn{\FMain{}}
		{
			Initialize $\mathbf{\Lambda}$ to a random integer vector: $0\le\lambda_i\le N, i\in\{1,2,\dots,4A_\text{max}^2\}$\;
			${\mathbf{\Pi}}\gets \FCalcSSD\left(\mathbf{\Lambda}\right)$\;
			\While{True}{
				$\mathbf{\Lambda}_\text{old}\gets\mathbf{\Lambda}$\;
				\For{$i\in\{1,2,\dots,4A_\text{max}^2\}$}{
					$\lambda_i\gets\arg\min\limits_\lambda\FPnt(\lambda,\mathbf{s}_i,{\mathbf{\Pi}})$\;
				}
				$\mathbf{\Pi}\gets \FCalcSSD\left(\mathbf{\Lambda}\right)$\;
				$\epsilon\gets2\sqrt{\frac{\left\Vert\mathbf{\Lambda}-\mathbf{\Lambda}_{\text{old}}\right\Vert}{\left\Vert\mathbf{\Lambda}+\mathbf{\Lambda}_{\text{old}}\right\Vert}}$\;
				\If{$\epsilon\le\epsilon_{\text{cvg}}$}{Break}\;
			}
		}
		\Return $\mathbf{\Lambda}$\;
		\;
		\Pn{\FCalcSSD{$\mathbf{\Lambda}$}}{
			$\mathbf{P}\gets\mathbf{0}_{4A_\text{max}^2\times4A_\text{max}^2}$\;
			\For{$i\in\{1,2,\dots,4A_\text{max}^2\}$}{
				\For{$j\in\{1,2,\dots,4A_\text{max}^2\}$}{
					$P_{i,j}\gets\FPTr(\lambda,\mathbf{s}_i,\mathbf{s}_j)$
				}
			}
			Solve $\mathbf{\Pi}$ from $\mathbf{P}$ according to Eq. \eqref{eq:steady-state_dist}
		}\Return $\mathbf{\Pi}$\;
		\;
		\Pn{\FPTr{$\lambda,\phi_k,\phi_{k+1}$}}{
			$n_{k,1}\gets\lambda,\quad n_{k,2}\gets(N-\lambda)$\;
			\For{$m\in\{1,2\}$}{
				Assign $\gamma_{k,m}$ according to Eq.~\eqref{eq:bernoulli_channel} and $\phi_k$\;
				Calculate $V_{k,m}$ and $C_{k,m}$ according to Eqs.~\eqref{eq:dispersion}, \eqref{eq:capacity}\;
				Calculate $\varepsilon_{k,m}(\phi_k)$ according to Eq.~\eqref{eq:err_rate}\;
			}
			Calculate $\text{Prob}(\phi_{k+1}\vert\phi_{k})$ according to \eqref{eq:transition}
		}\Return $\text{Prob}(\phi_{k+1}\vert\phi_{k})$\;
		\;
		\Pn{\FPnt{$\lambda,\mathbf{s}_i,\mathbf{\Pi}$}}{
		}
		\Return $\sum\limits_{i}\sum\limits_{j:\mathbf{s}_j\in\mathfrak{S}_{\text{out}}}\pi_i\FPTr(\lambda,\mathbf{s}_i,\mathbf{s}_j)$ \;
	\end{algorithm}
%\end{alg}
\caption{The recursive policy optimizer to minimize AoI outage}
\end{figure}

\subsection{\highlight{Heuristic Penalties against Immature Convergence}}\label{subsec:alternative_cost_functions}
\highlight{It shall be noted that the designed recursive framework relies on two implicit assumptions to converge to the global optimum, that over every iteration there shall be: \emph{i}) a strongly limited update in ${\mathbf{\Pi}\times\mathbf{P}}$, and \emph{ii}) an observable reduction in the penalty function. While the earlier one can be mostly satisfied\footnote{Or forced by artificially limiting the change-per-iteration if necessary}, the latter one is very likely challenged due to the sparsity of simple binary penalty \eqref{eq:simple_reward}, especially when the outage rate is low. Such problem of sparse reward has been widely observed in different applications, and known to bring the risk of immature convergence at local minimums, which may significantly lower the optimization performance \cite{RHL+2018learning}. Concerning of this, we propose several alternative heuristic penalty functions, which are more risk-sensitive in comparison to the simple binary penalty \eqref{eq:simple_reward}.}

\subsubsection{Mean Static Sum AoI}
By replacing the return value of \texttt{Pnt} with
\begin{equation}
	\sum\limits_{i,j}(A_{j,1}+A_{j,2})\pi_i\texttt{PTr}(\lambda,\mathbf{s}_i,\mathbf{s}_j),\label{eq:pi_cost_sum_aoi}
\end{equation}
the penalty is set to the mean sum AoI of two devices over the long-term state distribution.
% Since the outage event is determined by the maximal AoI between the two devices, this cost function is unlikely to minimize the outage rate, but only designed for the purpose of benchmarking.

\subsubsection{Mean Static Peak AoI}
An improvement can be made by using the expectation of peak AoI over the long-term state distribution, which is strongly correlated to the outage rate:
\begin{equation}
	\sum\limits_{i,j}\max\{A_{j,1},A_{j,2}\}\pi_i\texttt{PTr}(\lambda,\mathbf{s}_i,\mathbf{s}_j).
%\begin{split}
%\sum\limits_{j}\sum\limits_{i:\mathbf{s}_i\in\mathfrak{S}_{\text{out}}}&\left[\check{\mathbf{P}}_{i,j}\times\text{Prob}(\phi_{k+1}=\mathbf{s}_j\vert\phi_{k}=\mathbf{s}_\text{current})\right.\\&\left.\times\max\{A_{j,1},A_{j,2}\}\right].
%\end{split}
\label{eq:pi_cost_risk_linear}
\end{equation}

\subsubsection{Exponential Mean Static Peak AoI}
Concerning that the instantaneous risk of outage can dramatically (probably more  than linearly) increase along with the peak device AoI, we can follow the idea of risk-sensitiveness proposed in \cite{GSKN2018risk} to further modify the penalty function \eqref{eq:pi_cost_risk_linear} into an exponential variable
\begin{equation}
\sum\limits_{i,j}e^{\max\{A_{j,1},A_{j,2}\}}\pi_i\texttt{PTr}(\lambda,\mathbf{s}_i,\mathbf{s}_j).
%\begin{split}
%\sum\limits_{j}\sum\limits_{i:\mathbf{s}_i\in\mathfrak{S}_{\text{out}}}&\left[\check{\mathbf{P}}_{i,j}\times\text{Prob}(\phi_{k+1}=\mathbf{s}_j\vert\phi_{k}=\mathbf{s}_\text{current})\right.\\&\left.\times e^{\max\{A_{j,1},A_{j,2}\}}\right].
%\end{split}
\label{eq:pi_cost_risk_exp}
\end{equation}

\section{Outage Burstiness Analysis}\label{sec:burstiness}
Practically, besides the system outage rate, the burstiness of outage events can be also worth a discussion. More specifically, under the same outage rate, upon the system resilience, either sparse-and-long outage bursts or dense-but-short outage impulses may be preferred. For example, in sensitive and tolerance-critical systems that require manual operations to recover from every outage, the earlier case will cause less repair cost. In contrast, for the systems that are robust enough to overcome short outages or even autonomously recover from them, the latter case is obviously better. In the context of AoI outage, the burstiness is especially interesting in the FBL regime, where the packet error rate of every individual transmission is significantly higher than that in the infinite blocklength regime, and the occurrence of continuous outages can be therefore common enough to impact.

To analyze the burstiness of outages in the system under our investigation, we adopt the classical approach of \emph{Zorzi}, which was initially proposed to analyze the outage and error events in bursty channels~\cite{Zorzi1998outage}.

First, for the convenience of notation we denote by $P_{ij}=\text{Prob}\left(\left.\phi_k = s_j \right\vert \phi_{k-1} =s_i\right)$ the transition probability from $s_i$ to $s_j$, which can be obtained by~\eqref{eq:transition}.
%If the frame structure is fixed and channel is constant over the transmission. we have 
    % \begin{equation}
    %     P^m_{ij}=\left\{
    %                 \begin{array}{ll}
    %                 1-\varepsilon,\ {\rm if}\ j=1\\
    %                 \varepsilon, \ {\rm if}\ j=i+1\\
    %                 0, \ {\rm otherwise} 
    %                 \end{array}
    %                 \right.
    % \end{equation}
Then we further introduce the function $\xi_{ij}(k)$ to describe the probability that the system, starting from an arbitrary period $n$, undergoes a continuous outage $\left(\phi\in\mathfrak{S}_\text{out}\right)$ over the periods $n+1$, $n+2$, ..., $n+k-1$, and thereafter comes to the state $\mathbf{s}_j$ in the period $n+k$, which can be calculated as:
\begin{equation}
    \xi_{ij}(k)=\begin{cases}
    	P_{ij}&k=1,\\
    	\sum\limits_{l: \mathbf{s}_l\in\mathfrak{S}_\text{out}}\xi_{lj}(k-1)P_{il}&k\ge2.
    \end{cases}
%    \left\{
%                \begin{array}{ll}
%                \underset{s_m\in \mathfrak{S}_{\rm out}}{\sum} \xi_{mj}(n-1)P_{im},\ {\rm if}\ n\geq 2\\
%                P_{ij},\ {\rm if}\ n=1
%                \end{array}
%    \right.
\end{equation}
Obviously, the value of any $\xi_{ij}(k)$ can be iterated from $P_{ij}(1)$. 

Furthermore, let $\mathfrak{S}_\text{res}=\mathfrak{S}-\mathfrak{S}_\text{out}$ indicate the set of outage-free states, for all $\left(\mathfrak{A},\mathfrak{B}\right)\in \{\mathfrak{S}_\text{out},\mathfrak{S}_\text{res}\}^2$ we define
\begin{equation}
	\begin{split}
		\xi_{\mathfrak{A}\mathfrak{B}}(k)=\text{Prob}\left(\phi_n\in\mathfrak{A},\phi_{n+k}\in\mathfrak{B},\phi_l\in\mathfrak{S}_\text{out}\right),\\
		\forall n\in\mathbb{N},\forall k\in\mathbb{N},\forall n<l<n+k.
	\end{split}
\end{equation}
It can be then easily derived that
%Furthermore, we define $\mathfrak{S}_\text{out}$ the set of all states with $A_m(k)>A_{\rm out}$ and $\mathfrak{S}_\text{res}$ the set of rest states. Then, we define $\xi_{\mathcal{R}\mathcal{S}}$, where $\mathcal{R},\mathcal{S}\in \{\mathfrak{S}_\text{out},\mathfrak{S}_\text{res}\}$, as the probability that $A(k)$ is in $\mathcal{R}$ at time $n$, in $\mathcal{S}$ at time $n+k$ and in $\mathfrak{S}_\text{out}$ in between, which can be written as
\begin{equation}
	\xi_{\mathfrak{A}\mathfrak{B}}(k)=\sum_{i:\mathbf{s}_i\in\mathfrak{A}}\sum_{j:\mathbf{s}_j\in\mathfrak{B}}\pi_i\xi_{ij}(k).
%    \xi_{\mathcal{R}\mathcal{S}}(n)=\sum_{i\in\mathcal{R}}\sum_{j\in\mathcal{S}}\pi_i\xi_{ij}(n),
\end{equation}
%where $\pi_i$ is the steady-state probability of being in state $\mathbf{s}_i$. Since we have assumed in Sec.~\ref{sec:model} that $\phi_0=\mathbf{s}_1=(1,1,0,0)$, there is $\pi_i=\check{\mathbf{P}}_{1,i}$.

Now we denote by $T_\text{out}$ the random process of outage duration. More specifically, $T_\text{out}=t$ indicates an event that the system state starts in $\mathfrak{S}_\text{res}$, then undergoes the following $t-1$ periods in $\mathfrak{S}_\text{out}$ and ends again in $\mathfrak{S}_\text{res}$. Thus,
\begin{equation}
    P_{T_\text{out}}(t)=\frac{\xi_{\mathfrak{S}_\text{res}\mathfrak{S}_\text{res}}(t+1)}{\xi_{\mathfrak{S}_\text{res}\mathfrak{S}_\text{out}}(1)},
\end{equation}
and we can obtain the mean duration of outages $T_\text{out}$:
\begin{equation}
	\mathbb{E}\left(T_\text{out}\right)=\sum\limits_{t\ge1}tP_{T_\text{out}}(t)=1+\sum_{t=2}^{+\infty}\frac{\xi_{\mathfrak{S}_\text{res}\mathfrak{S}_\text{out}}(t)}{\xi_{\mathfrak{S}_\text{res}\mathfrak{S}_\text{out}}(1)}.\label{eq:mean_outage_duration}
\end{equation}
%and
%\begin{equation}
%	\begin{split}
%	    &\text{Var}\left(T_\text{out}\right)=\mathbb{E}\left\{\left[T_\text{out}-\mathbb{E}\left(T_\text{out}\right)\right]^2\right\}\\
%	  	=&\mathbb{E}\left(T_\text{out}^2\right)-\mathbb{E}\left(T_\text{out}\right)^2=\sum_{t=1} t^2P_{T_\text{out}}(t)-\mathbb{E}\left(T_\text{out}\right)^2\\
%	 	=&1+3\sum_{t\ge2}\frac{\xi_{\mathfrak{S}_\text{res}\mathfrak{S}_\text{out}}(t)}{\xi_{\mathfrak{S}_\text{res}\mathfrak{S}_\text{out}}(1)}-\mathbb{E}\left(T_\text{out}\right)^2,
%	\end{split}
%\end{equation}
%respectively.

Note that with Eq.~\eqref{eq:mean_outage_duration}, the outage probability $P_\text{out}$ defined in Eq.~\eqref{eq:outage_prob_from_transition} can now also be represented as:
\begin{equation}
    P_\text{out}=\xi_{\mathfrak{S}_\text{res}\mathfrak{S}_\text{out}}(1)\mathbb{E}\left(T_\text{out}\right).
\end{equation}

Similarly, the expectation of non-outage duration, or inter-outage interval (IoI), $T_\text{res}$ can be estimated as:
\begin{equation}
    \mathbb{E}\left(T_\text{res}\right)=\frac{1-P_\text{out}}{\xi_{\mathfrak{S}_\text{res}\mathfrak{S}_\text{out}}(1)}.\label{eq:mean_ioi}
\end{equation}

\section{Numerical Evaluations}\label{sec:sim}
\subsection{Setup of the Simulation Companion}
To validate our analyses and proposed optimization approach, we implemented a simulation companion in \emph{Julia}~\cite{BEKS2017julia}. The system model as predefined in Sec.~\ref{sec:model} is configured to the specifications listed in Tab.~\ref{tab:spec}. Specifically, three Bernoulli fading profiles are defined, to present the different scenarios where: \emph{A}) both channels have slight fading effects; \emph{B}) both channels undergo significant fadings; and \emph{C}) the fading conditions are polarized for the two sensors, respectively.
\begin{table}[!hbtp]
	\centering
	\caption{Specifications of important simulation parameters}
	\label{tab:spec}
	\begin{tabular}{l|c|c}
		\toprule[2px]
		\textbf{Category}&\textbf{Parameter}&\textbf{Value}\\
		\midrule[1.5px]
		%\multirow{6}{*}{Channel}&$B$&\SI{1}{\hertz}\\\cline{3-3}
		\multirow{5}{*}{Channel}&&Scenario A: $(0.9,0.7)$\\
		&$(\alpha_1,\alpha_2)$&Scenario B: $(0.6,0.4)$\\
		&&Scenario C: $(0.9,0.2)$\\%\cline{3-3}
		&&\\
		&$(\gamma_\text{good},\gamma_\text{bad})$&$(\SI{-12.2}{\dB},\SI{-15.2}{\dB})$\\
		\midrule[1px]
		\multirow{2}{*}{Blocklength}&$N$&$\SI{1000}{\bit}$\\
		&$d$&$\SI{16}{\bit}$\\
		\midrule[1px]
		\multirow{3}{*}{System state}&$A_\text{max}$&5\\
		&$A_\text{out}$&3\\
		&$\phi_0$&$(1,1,0,0)$\\
		\midrule[1px]
		Optimizer convergence &$\epsilon_{\text{cvg}}$ &$1\times10^{-5}$\\
		\bottomrule[2px]
	\end{tabular}
\end{table}

\subsection{Evaluation of Blocklength Allocation Optimizers}
\highlight{To evaluate the performance of our proposed recursive policy  optimizer, we apply it to calculate the optimal blocklength allocation policies (denoted by ``optimized'') in all three scenarios, both with the original penalty function \texttt{Pnt} and all the alternative heuristic penalties introduced in Sec.~\ref{subsec:alternative_cost_functions}. As benchmarks, we also consider two other reference policies in addition: one is to naively share the blocklength between both devices, i.e. $n_{k,1}=n_{k,2}=\frac{N}{2},\forall k$; the other is to allocate the blocklength at every step $k$ to minimize the sum transmission error $(\varepsilon_{k,1}+\varepsilon_{k,2})$, which is a state-of-the-art solution in FBL regime~\cite{Poly_2010_FBL}. We evaluate every policy through $100$ repetitions of Monte-Carlo test, where each individual test simulation lasts $2500$ periods. The results are listed in Tab.~\ref{tab:sim_outage_rate}.
	\begin{table}
		\centering
		\caption{Measured AoI outage rates with different blocklength allocation policies and reference scenarios}
		\begin{tabular}{l|l|c|c|c}
			\toprule[2px]
			\multicolumn{2}{l|}{\diagbox{\textbf{Policy}}{\textbf{Scenario}}}&{A}&{B}&{C}\\
			\midrule[1.5px]
			\multirow{4}{*}{Optimized}& (binary penalty)			&0.91\%&3.31\%&1.37\%\\
			& (min. sum AoI)			&0.28\%&2.08\%&1.32\%\\
			& (min. peak AoI)			&0.29\%&1.68\%&1.25\%\\
			& (min. exp. peak AoI)	&0.25\%&1.39\%&1.20\%\\\midrule[1.5px]
			\multirow{2}{*}{Benchmarks}&naive equal sharing&0.73\%&3.25\%&3.94\%\\
			&min. error rate		&0.26\%&1.65\%&1.21\%\\
			\bottomrule[2px]
		\end{tabular}
		\label{tab:sim_outage_rate}
\end{table}}

\highlight{It can be observed that the proposed method fails to find the global optimum with the simple binary penalty, resulting in a performance poorer than that of the single-step error minimization policy, as we have concerned due to the sparsity of penalty. In contrast, when enhanced with the heuristic penalties, the optimizer becomes capable to outperform conventional benchmarks. Especially, the exponential static peak AoI appears to be the most outstanding penalty function.}

\subsection{Verification of Burstiness Analysis}
Then, to validate our analysis to the burstiness of AoI outage under arbitrary blocklength allocation policy, we conduct a Monte-Carlo test with $100$ random policies. We apply each policy to the system in Scenario B, simulate the system for a duration of $10 000$ periods, and record the AoI outage events. Then we observe the recorded outage history at the time instants of $500$, $1000$, $2500$, $5000$, and $10 000$ periods, respectively. With each observation, we measure the AoI outage rate $P_\text{out}$, the mean outage duration $\overline{T}_\text{out}$, and the mean IoI $\overline{T}_\text{res}$, and compare them to the values estimated according our analyses \eqref{eq:outage_prob_from_ssd}, \eqref{eq:mean_outage_duration}, and \eqref{eq:mean_ioi}, respectively. 

\begin{figure}[!hbtp]
	\centering
	\includegraphics[width=.9\linewidth]{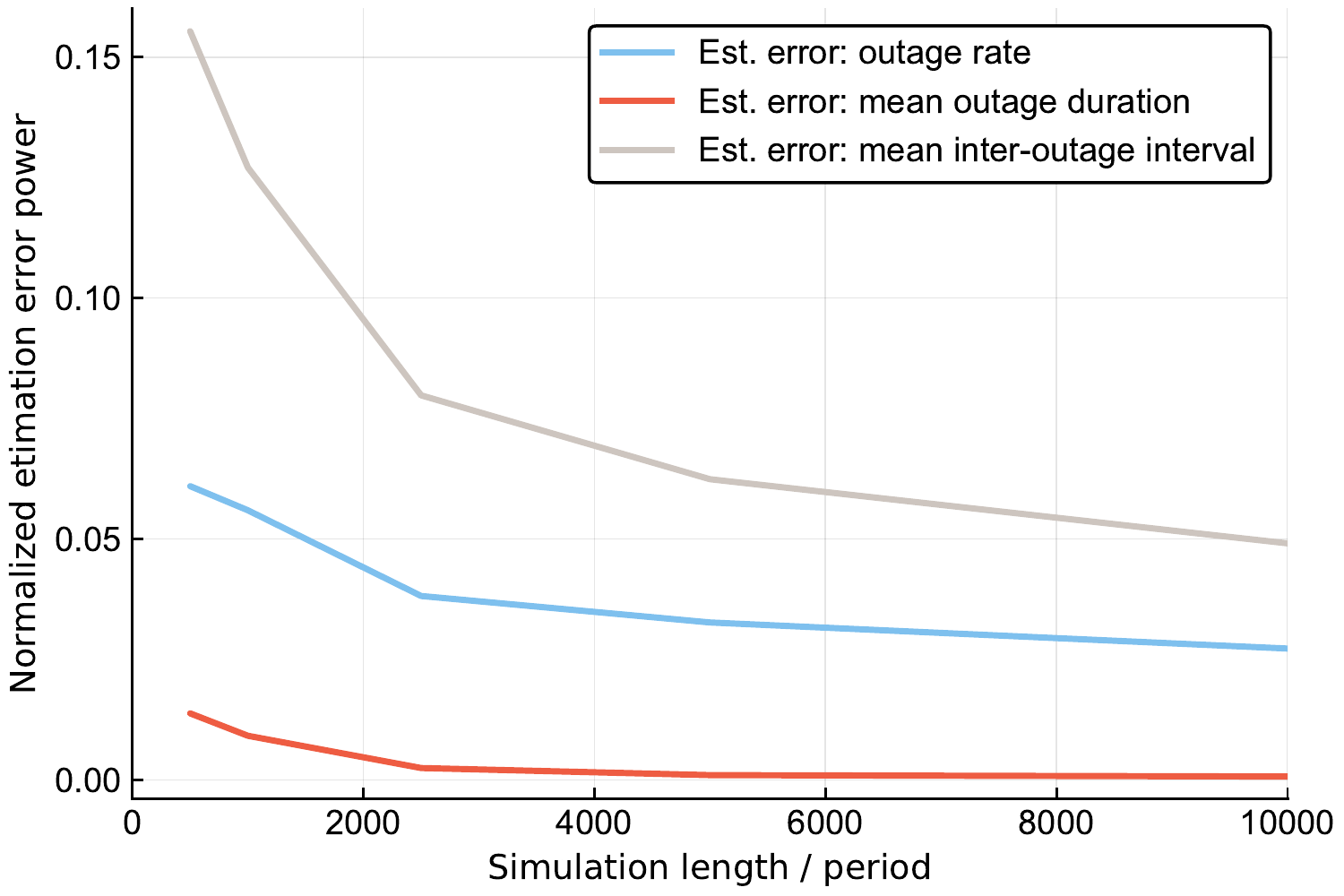}
	\caption{Measured statistics converge towards the static-state-based estimations along with increasing simulation length.}
	\label{fig:burstiness_analysis}
\end{figure}

We calculate the estimation errors in their powers normalized to the measured value, which are illustrated in Fig.~\ref{fig:burstiness_analysis}. It can be seen that the errors between measurement and estimation are relatively significant in short term, \highlight{but rapidly converge to negligible levels ($<5\%$) as the simulation continues}. This verifies the accuracy of our analyses, while the errors at beginning of simulation is caused by the variance of observations, which is considerably high due to the sparsity of outage events (which also implies small sample spaces of outage durations and inter-outage intervals).

We would like to note that the burstiness analysis is independent from our proposed policy optimization approach, i.e. the bursty feature can be estimated on any given blocklength allocation policy, and are thereby available as secondary objective functions for policy optimization upon system demand.

\section{Conclusion and Outlooks}\label{sec:concl}
\highlight{In this paper, we have studied the phenomenon of AoI outage in OMA systems as a blocklength allocation problem. We have proven that the AoI outage rate does not rely on the converging path of system state, but only its static distribution. Therewith, we have proposed a recursive policy optimizer with heuristic penalty functions to enhance its convergence. The proposed approach has been numerically demonstrated as effective. Furthermore, in the context of finite blocklength, we have discussed the bursty behavior of outage events, and proposed to accurately estimate the outage burstiness under any certain policy.}

\highlight{We see rich potentials in this work to be further deepened and extended. First, concerning the time complexity as high as $\mathcal{O}\left(\vert\mathfrak{S}\vert^3\right)$ in exhaustive iteration, it appears essential to replace it with a Reinforcement Learning solver for better  scalability. Second, our long-term state analysis on this problem actually reveals a more generic conclusion, that the reliability of ergodic Markov systems, as a long-term statistical feature, relies only on the long-term state, which can be generalized for many other robustness-critical applications in a wide range. Furthermore, we consider it feasible to invoke our proposed outage burstiness estimators in the policy optimization for a more advanced control on the high-order system dynamics. Additionally, it is worth to test our proposed approach under more detailed and practical system specifications, to evaluate its capability to fulfill certain B5G/6G requirements.
}
% Can use something like this to put references on a page
% by themselves when using endfloat and the captionsoff option.
\ifCLASSOPTIONcaptionsoff
  \newpage
\fi

% that's all folks

\begin{thebibliography}{10}


\bibitem{KYG2012real}
S. Kaul, R. Yates, and M. Gruteser, ``Real-time status: How often should one update?'' in \emph{Proc. IEEE Conf. Comput. Commun. (INFOCOM)}, pp. 2731–2735, Mar. 2012.

\bibitem{HLS+2019real}
K. Huang, W. Liu, M. Shirvanimoghaddam, et al., ``Real-time remote estimation with Hybrid ARQ in wireless networked control,'' arXiv preprint arXiv:1903.12472.

\bibitem{Hsu2018age}
Y.-P. Hsu, ``Age of information: Whittle index for scheduling stochastic arrivals,'' in \emph{Proc. IEEE Int. Symp. Inf. Theor. (ISIT)}, pp. 2634-2638. IEEE, Jun. 2018.

\bibitem{KSU+2018scheduling}
I. Kadota, A. Sinha, E. Uysal-Biyikoglu, et al., ``Scheduling policies for minimizing age of information in broadcast wireless networks,'' in \emph{IEEE/ACM Trans. on Networking}, vol. 26, pp. 2637-2650. Dec. 2018.

\bibitem{JZNC2019unified}
Z. Jiang, S. Zhou, Z. Niu, and Y. Cheng, ``A unified sampling and scheduling approach for status update in wireless multiaccess networks,'' in \emph{Proc. IEEE Conf. Comput. Commun.  (INFOCOM)}, pp. 208--216. IEEE, Apr. 2019.

\bibitem{JKZ+2018decentralized}
Z. Jiang, B. Krishnamachari, X. Zheng, et al., ``Decentralized status update for age-of-information optimization in wireless multiaccess channels,'' in \emph{Proc. IEEE Int. Symp. Inf. Theor. (ISIT)}, pp. 2276--2280. IEEE, Jun. 2018.

\bibitem{JKZN2018can}
Z. Jiang, B. Krishnamachari, S. Zhou, and Z. Niu, ``Can decentralized status update achieve universally near-optimal age-of-information in wireless multiaccess channels?,'' in \emph{Proc. IEEE Int. Symp. Inf. Theor. (ISIT)}, pp. 144--152. IEEE, Jun. 2018.

\bibitem{HZJ+2019optimal}
Bin Han, Yao Zhu, Zhiyuan Jiang, et al., ``Optimal blocklength allocation towards reduced age of information in wireless sensor networks,'' in \emph{IEEE GLOBECOM 2019 Workshop on Wireless Edge Intelligence}, Waikoloa, HI, USA, Dec. 2019.

\bibitem{ITU2017minimum}
``Minimum requirements related to technical performance for IMT–2020 radio interface(s)'', document ITU-R M.2410-0, International Telecommunication Union-Recommendations, Nov. 2017.

\bibitem{3GPP2017tr38802}
``Study on new radio (NR) access technology physical layer aspects'', documment TR 38.802, 3GPP, Mar. 2017.

\bibitem{PKBF2017rethink}
K. I. Pedersen, S. R. Khosravirad, G. Berardinelli, and F. Frederiksen, ``Rethink hybrid automatic repeat request design for 5G: Five configurable enhancements,'' in \emph{IEEE Wireless Communications}, vol. 24, no. 6, pp. 154--160, Dec. 2017.

\bibitem{EBDG2019radio}
S. E. Elayoubi, P. Brown, M. Deghel and A. Galindo-Serrano, ``Radio resource allocation and retransmission schemes for URLLC over 5G networks,'' in \emph{IEEE Journal on Selected Areas in Communications}, vol. 37, no. 4, pp. 896--904, Apr. 2019.

\bibitem{KNBP2018uplink}
R. Kotaba, C. Navarro Manch\'on, T. Balercia and P. Popovski, ``Uplink Transmissions in URLLC Systems With Shared Diversity Resources,'' in \emph{IEEE Wireless Communications Letters}, vol. 7, no. 4, pp. 590--593, Aug. 2018.

\bibitem{RPD+2019resource}
H. Ren, C. Pan, Y. Deng, et al., ``Resource Allocation for URLLC in 5G Mission-Critical IoT Networks,'' in \emph{Proc. 2019 IEEE Int. Conf. Commun. (ICC)}, Shanghai, China, 2019, pp.1--6.

\bibitem{GSKN2018risk}
Xueying Guo, Rahul Singh, P. R. Kumar, and Zhisheng Niu, ``A risk-sensitive approach for packet inter-delivery time optimization in networked cyber-physical systems,'' in \emph{IEEE/ACM Transactions on Networking (TON)}, vol. 26, no. 4, pp. 1976--1989. IEEE, 2018.

\bibitem{Poly_2010_FBL}
Y. Polyanskiy, H. V. Poor, and S. Verd\'u, “Channel coding rate in the finite blocklength regime,” \emph{IEEE Trans. Inf. Theory}, vol. 56, no. 5, pp. 2307–2359, May 2010.

\bibitem{Khan_2017_FBL}
T. A. Khan, R. W. Heath, Jr., and P. Popovski, “Wirelessly powered communication networks with short packets,” in \emph{IEEE Trans. Commun.}, vol. 65, no. 12, pp. 5529--5543, Dec. 2017.

\bibitem{Wang_2019_FBL}
H. Wang, Q. Yang, Z. Ding and H. V. Poor, ``Secure short-packet communications for mission-critical IoT applications," in \emph{IEEE Trans. Commun.}, vol. 18, no. 5, pp. 2565--2578, May 2019.

\bibitem{Hu_2019_FBL}
Y. Hu, Y. Zhu, M. C. Gursoy and A. Schmeink, ``SWIPT-Enabled Relaying in IoT Networks Operating With Finite Blocklength Codes," in \emph{IEEE J. Sel. Areas Commun. (JSAC)}, vol. 37, no. 1, pp. 74--88, Jan. 2019.

\bibitem{Zhu_2019_FBL}
Y. Zhu, Y. Hu, A. Schmeink and J. Gross, "Energy minimization of mobile edge computing networks with finite retransmissions in the finite blocklength regime," \emph{2019 IEEE 20th International Workshop on Signal Processing Advances in Wireless Communications (SPAWC)}, Cannes, France, 2019, pp. 1--5.

\bibitem{Avranas_2019_FBL}
A. Avranas, M. Kountouris and P. Ciblat, "Energy-latency tradeoff in ultra-reliable low-latency communication with retransmissions," in \emph{IEEE J. Sel. Areas Commun. (JSAC)},  vol. 36,  no. 11, pp. 2475-2485, Nov. 2018.

\bibitem{RHL+2018learning}
M. Riedmiller, R. Hafner, T. Lampe, et al., ``Learning by playing - Solving sparse reward tasks from scratch,'' arXiv preprint arXiv:1802.10567. Feb. 2018.

\bibitem{KR2000policy}
D. Koller, and P. Ronald, ``Policy iteration for factored MDPs,'' in \emph{Proc. 16th Conf. Uncertainty in Artificial Intelligence}, pp. 326--334. Morgan Kaufmann Publishers Inc., 2000.

\bibitem{Zorzi1998outage}
M. Zorzi, ``Outage and error events in bursty channels,'' in \emph{IEEE Trans. Commun.}, vol. 46, no. 3, pp. 349--356, Mar. 1998.

\bibitem{BEKS2017julia}
Jeff Bezanson, Alan Edelman, Stefan Karpinski, and Viral B. Shah, ``Julia: A fresh approach to numerical computing,'' in \emph{SIAM Review}, vol. 59, no.1, pp. 65--98, Feb. 2017.
\end{thebibliography}
\end{document}